
\input phyzzx
\sequentialequations

\hsize=6.5in
\vsize=8.7in

\overfullrule=0pt
\catcode`\@=11
\def\mm{matrix model}

\def \CO{{\cal O}}

\def\npb {Nucl. Phys.}
\def\plb {Phys. Lett.}
\def\prd {Phys. Rev.}

\def\mpla{Mod. Phys. Lett.}
\def\cmp {Commun. Math. Phys.}
\def\bb#1       {{{\tt hep-th/#1}}}

\def\td{two-dimensional}
\def\opf{odd-point functions}

\def\eqaligntwo#1{\null\,\vcenter{\openup\jot\m@th
\ialign{\strut\hfil
$\displaystyle{##}$&$\displaystyle{{}##}$&$\displaystyle{{}##}$\hfil
\crcr#1\crcr}}\,}
\catcode`\@=12

\font\cmss=cmss10 \font\cmsss=cmss10 at 7pt
\def\IZ{\relax\ifmmode\mathchoice
   {\hbox{\cmss Z\kern-.4em Z}}{\hbox{\cmss Z\kern-.4em Z}}
   {\lower.9pt\hbox{\cmsss Z\kern-.4em Z}}
   {\lower1.2pt\hbox{\cmsss Z\kern-.4em Z}}\else{\cmss Z\kern-.4em Z}\fi}
\def\dj{\hbox{d\kern-0.347em \vrule width 0.3em height 1.252ex depth
-1.21ex \kern 0.051em}}

\def\half{{1\over 2}}

\REF\rWi{E. Witten, \prd\ {\bf D44}, 314 (1991).}
\REF\rDVV {R. Dijkgraaf, E. Verlinde and H. Verlinde, \npb\ {\bf B371}, 269
(1992).}
\REF\rBH{G. Mandal, A. Sengupta and S. Wadia, \mpla\ {\bf A6}, 1685 (1991);
I. Bars and D. Nemeschansky, \npb\ {\bf B348}, 89 (1991);
S. Elizur, A. Forge and E. Rabinovici, \npb\ {\bf B359}, 581 (1991);
M. Ro\v cek, K. Schoutens and A. Sevrin, \plb\ {\bf B265}, 303 (1991).}
\REF\rBHmm{
S.~R. Das, \mpla\ {\bf A8}, 69 (1993);
A. Dhar, G. Mandal and S. Wadia, \mpla\ {\bf A7}, 3703 (1992);
J.~G. Russo, \plb\ {\bf B300}, 336 (1993);
Z. Yang, {\it ``A possible black hole background in $c=1$ matrix
model''}, UR-1251, \bb{9202078}.}
\REF\rJeYo{A. Jevicki and T. Yoneya, {\it ``A deformed matrix model and
the black hole background in two-dimensional string theory''},
NSF-ITP-93-67, BROWN-HEP-904, UT-KOMABA/93-10, \bb{9305109}.}
\REF\rAvJe{J. Avan and A. Jevicki, \cmp\ {\bf 150}, 149 (1992).}
\REF\rCFT{M. Bershadsky and D. Kutasov, \plb\ {\bf B266}, 345 (1991);
E. Martinec and S. Shatashvili, \npb\ {\bf B368}, 338 (1992).}
\REF\rEg{T. Eguchi, {\it ``$c=1$ Liouville theory perturbed by the
black-hole mass operator''}, UT 650, \bb{9307185}.}
\REF\rDeRo{K. Demeterfi and J.~P. Rodrigues, {\it ``States and
quantum effects in the collective field theory of a deformed matrix model''},
PUPT-1407, CNLS-93-06, \bb{9306141}.}
\REF\rDa{U. Danielsson, {\it ``A matrix-model black hole''},
CERN-TH.6916/93, \bb{9306063}.}
\REF\rMPR{G. Moore, R. Plesser and S. Ramgoolam, \npb\ {\bf 377}, 143
(1992).}
\REF\rI{ I. R. Klebanov, {\it ``String theory in two dimensions''},
in ``String Theory and Quantum Gravity'',  Proceedings of the Trieste
Spring School 1991, eds. J. Harvey et al., (World Scientific,
Singapore, 1992).}
\REF\rGrKl{D.~J. Gross, I.~R. Klebanov and M.~Newman, \npb\ {\bf 350},
621 (1991);
D.~J. Gross and I.~R. Klebanov, \npb\ {\bf B359}, 3 (1991).}
\REF\rDiFrKu{P. Di Francesco and D. Kutasov, \plb\ {\bf B261}, 385 (1991).}

\nopagenumbers

{\baselineskip=16pt
\line{\hfil PUPT-1416}
\line{\hfil CNLS-93-09}
\line{\hfil {\tt hep-th/9308036}}
 }

\bigskip
\title{{\bf The Exact $S$-matrix of the Deformed $c=1$ Matrix Model }}
\bigskip

\centerline {{\caps Kre\v simir Demeterfi}\foot{{\rm On leave of absence
from the Ru\dj er Bo\v skovi\'c Institute, Zagreb, Croatia}}
and {\caps Igor R. Klebanov} }
\centerline{\sl Joseph Henry Laboratories}
\centerline{\sl Princeton University}
\centerline{\sl Princeton, New Jersey 08540}
\medskip
\centerline{and}
\medskip
\centerline{{\caps Jo\~ao P. Rodrigues}}
\centerline{{\sl Physics Department and Centre for Nonlinear Studies}}
\centerline{{\sl University of the Witwatersrand}}
\centerline{{\sl Wits 2050, South Africa}}

\bigskip
\abstract
We consider the $c=1$ \mm\ deformed by the operator $\half
M\Tr\Phi^{-2}$, which was conjectured by Jevicki and Yoneya to
describe a \td\ black hole of mass $M$. We calculate the exact
non-perturbative $S$-matrix and show that all the amplitudes involving
an odd number of particles vanish at least to all orders of perturbation
theory. We conjecture that these amplitudes vanish non-perturbatively
and prove this for the $2n \to 1$ scattering. For the 2-- and 4--particle
amplitudes we give some leading terms of the perturbative expansion.
\vfill
\line{8/93\hfill}
\endpage

\pagenumbers

There has been a considerable amount of speculation on the
relation between the $c=1$ matrix model and two-dimensional stringy
black holes [\rWi,\rDVV,\rBH,\rBHmm].
Recently Jevicki and Yoneya [\rJeYo] made an interesting proposal
that a stationary black hole of mass $M$ is described by the large-$N$
Hermitian matrix quantum mechanics with potential
$U(\Phi)=\half\Tr (- \Phi^2 +M\Phi^{-2})$. The matrix eigenvalues
act as free fermions, and their Fermi level $\mu$ is set to zero.
The deformation of the $c=1$ matrix model by the operator $\Tr \Phi^{-2}$
is uniquely determined by the requirement that it preserve the
$w_\infty$ symmetry structure [\rAvJe]. There are further arguments
why operators with negative powers of $\Phi$ should be identified
with ``wrongly dressed'' Liouville theory operators, of which the black
hole mass perturbation is the leading example [\rJeYo,\rCFT--\rDa].
In Ref.~[\rJeYo] some
calculations were performed in the deformed \mm\ with a number of
intriguing results. It was found that $1/\sqrt M$ plays the role
of the string coupling constant $g_{\rm st}$, in agreement with
string theory in the \td\ black hole background.
The tree level \opf\ were found to vanish, which provided
one more argument in favor of the black hole analogy.
Further studies of the deformed model, including some loop
corrections, were performed in [\rDeRo,\rDa].

In this Letter we calculate the exact non-perturbative $S$-matrix
of the fermion density perturbations in the
deformed \mm. We find that all the \opf\ vanish at least to all orders
in $g_{\rm st}$. Furthermore, we show that
the $2k\to 1$ amplitudes vanish non-perturbatively and conjecture that
this is true for  \opf\ with other kinematical structures.
For the 2-- and 4--point functions we give
a few leading terms of the loop expansion.

Our exact solution of the deformed matrix model is based on the powerful
method of Moore, Plesser and Ramgoolam [\rMPR], who constructed
the $S$-matrix of the $c=1$ \mm\
in terms of the single-fermion reflection coefficient.
Remarkably, in the deformed model the reflection coefficient can also
be calculated exactly. The crucial observation is that the single
fermion wave function with energy $(-\epsilon)$, which satisfies the
Schr\" odinger equation
$$ \left ({d^2\over dx^2} +x^2-{M\over x^2} -2\epsilon\right )
\psi_\epsilon (x)=0\ ,\eqno\eq $$
is explicitly given by
$$\psi_\epsilon (x)={1\over\sqrt{2\pi x}}\,
e^{-{i\pi\over 2}\,(\alpha+{1\over 2})}\,\,
e^{-\epsilon\pi/4}\,\,
{|\Gamma({1\over 2}+{i\epsilon\over 2}+\alpha)|\over
\Gamma(2\alpha+1)}\,\,
M_{i\epsilon/2,\alpha}(ix^2)\,\,,
\eqn\sol$$
where $\alpha={1\over 4}\sqrt {1+4 M}$ and $M_{i\epsilon/2,\alpha}$
is the Whittaker function.
The wave function is properly normalized
and satisfies the correct boundary condition $\psi_\epsilon(0)=0$.
The scattering phase shift can be read off from the asymptotic formula,
$$\psi_\epsilon (x\to \infty)=
{1\over\sqrt{2\pi x}}\,
\Bigl( e^{-ix^2/2}\,e^{i\epsilon\ln x}\,S +
e^{ix^2/2}\,e^{-i\epsilon\ln x}\,S^* \,\Bigr)\,\,,
\eqn\asol$$
where
$$S\equiv e^{i\pi(2\alpha +1)/4}\,\,
\sqrt{{\Gamma({1\over 2}-{i\epsilon\over 2}+\alpha)\over
\Gamma({1\over 2}+{i\epsilon\over 2}+\alpha) }}\,\,.$$
Now we can calculate the asymptotic behavior of the resolvent
$I(x_1, x_2) =\langle x_1 |{1\over H-\mu-iq}| x_2 \rangle$. Introducing the
classical time $\tau$ through
$x^2 (\tau)= \mu+\sqrt {M+\mu^2}\cosh (2\tau)$,
we find
$$\eqalign{
I(x_1,x_2;q>0) \mathrel{\mathop=_{x_1,x_2\to \infty}}
{i\over\sqrt{x_1 x_2}}\,&\Bigl\{
e^{i\, |\, G(\tau_1)-G(\tau_2)\,|}\,
e^{-q|\tau_1-\tau_2|} \cr
&+R_q\,e^{i (\, G(\tau_1)+G(\tau_2)\,)}\,
e^{-q(\tau_1+\tau_2)} \,\Bigr\}\,\,,}\eqn\res$$
where
$G(\tau)=-{1\over 4}\, \sqrt {M+\mu^2}\,e^{2\tau}+\mu\tau+\pi/4+
\CO(e^{-2\tau})$ is the
WKB phase factor for large $\tau$.  The reflection coefficient is
$$ R_q=\left ({4\over M+\mu^2}\right )^{|q|/2}
{\Gamma (\half -{i\mu\over 2}+ {|q|\over 2}+\alpha)\over
\Gamma (\half +{i\mu\over 2} - {|q|\over 2}+\alpha)} \,\,
e^{i\,\bigl[\,{1\over 2}\,\mu\,\log[(M+\mu^2)/4]-\mu+
\pi\alpha\,\bigr]}\,\,.
\eqn\refl$$
As shown in Ref.~[\rMPR], any scattering amplitude of the fermion
density perturbations
can be written in terms of integrals of products of reflection
coefficients.  Schematically, the relation is [\rMPR]
$$A(q_i)= \sum \int \prod \bigl (R_Q\,R_Q^*\bigr )\,\,.\eqn\scheme $$
The $l\to m$ amplitude is
$A_{l\to m}(q_1,\ldots, q_l; -q_{l+1}, \ldots, -q_{l+m})$,
where all $q_i$ are taken to be positive.
For now we work in the Euclidean domain and later
continue to the Minkowski signature.
The explicit formula for the $n\to 1$ amplitude reads
$$\eqalign{
A_{n\to 1}(&q_1,\ldots,q_n;-q)= i^{n+1}\,\Bigl\{\,\,
\sum_{\{i_1\}}\,\, \int_{q_{i_1}}^{q}\,\,dx \,R_{q-x}\,R_x^* -
\sum_{\{i_1,i_2\}}\,\,\int_{q_{i_1}+q_{i_2}}^{q}\,\,dx\,
R_{q-x}\,R_x^* +\cr
&\ldots + (-1)^{n-1}\,\sum_{\{i_1,\ldots,i_{n-1}\}} \,\,\,
\int_{q_{i_1}+\ldots +q_{i_{n-1}}}^{q}\,\,dx \,R_{q-x}\,R_x^*
- \int_0^{q}\,\,dx \,R_{q-x}\,R_x^*
\,\,\Bigr\}\,\,,}\eqn\genampl$$
where $\{i_1,i_2,\ldots ,i_k\}$ is a subset of $\{1,2,\ldots,n\}$.
Similarly the $2\to 2$ amplitude in the kinematic region
$q_1=\max \{q_i\}$ is given by
$$\eqalign{&A_{2\to 2} (q_1,q_2;-q_3,-q_4)=
-\int_{q_1}^{q_1+q_2} \,\,dx \,R_{q_1+q_2-x}\,R_x^*
-\int_{0}^{q_2}\,\,\,dx \,R_{q_1+q_2-x}\,R_x^* \cr
&+{1\over 2}\,\Bigl\{\,
\int_0^{q_2}\,\,dx\,R_{q_3-x}\,R_{q_2-x}\,R^*_{x+q_1-q_3}\,R^*_{x}
+\int_{q_3-q_2}^{q_3}\,\,dx\,R_{q_3-x}\,R_{q_1-x}\,
R^*_{x+q_2-q_3}\,R^*_{x} + \bigl( q_3 \to q_4\bigr)\,\Bigr\}\,\,.}
\eqn\twotwo$$

Our goal is to generate the asymptotic expansions of correlation
functions in powers of $g_{\rm st}=1/\sqrt M$.
In the following we set the Fermi level $\mu$ to zero, according to
the proposal of Ref.~[\rJeYo].
Our methods work equally well for $\mu\neq 0$, and we will
report those results in a later publication.
Let us first find the
asymptotic expansion of the reflection coefficient.
Introducing $r_q \equiv e^{-i\pi\alpha}\,R_q$, we have
$$\eqalign{&r_q= \Bigl(1+{1\over 4M}\Bigr)^{|q|/2} \,F(\alpha, q)\ ,\cr
\noalign{\vskip 0.2cm}
& F(\alpha, q)=\alpha^{-|q|}\,\,
{\Gamma (\half+ {|q|\over 2}+\alpha)\over
\Gamma (\half- {|q|\over 2}+\alpha)}\,\,. \cr } \eqn\rc$$
It is easy to show that
$$F(-\alpha, q)=F(\alpha, q)\,\,
{1+e^{-2\pi i\alpha} e^{-\pi i |q|}\over
1+e^{-2\pi i\alpha} e^{\pi i |q|}}\,\,. \eqn\sym$$
The fraction on the right-hand side is equal to 1, up to terms that
are invisible in the asymptotic expansion in powers of $1/\alpha$.
Therefore, the odd powers are absent from the asymptotic expansion,
$$ F(\alpha, q)= 1+ \sum_{k=1}^\infty d_k (q)\, \alpha^{-2k}\,\,.
\eqno\eq$$
It follows that there are no odd powers of $1/\sqrt M$ in the
asymptotic expansion of the reflection coefficient,
$$r_q (M)=1+ \sum_{k=1}^\infty c_k (q)\, M^{-k}\,\,.\eqn\asexp$$
The first few coefficients are given by
$$\eqalign{
&c_1 (q)={1\over 24}\,q(7-4q^2)\,\,, \cr
&c_2 (q)={1\over 5760}\,q(q-2)(501+128q-536q^2-128q^3+80q^4)\,\,,\cr
&c_3 (q)={1\over 2903040}\,q(q-2)(q-4)(115173+67968q-137060q^2\cr
&\qquad\qquad\qquad\qquad -78720q^3 +25072q^4+10752q^5-2240q^6)\,\,.}
\eqn\coeff$$
Simple scaling arguments indicate that all the \opf\ are expanded
in odd powers of $g_{st}=1/\sqrt M$. However, these powers are missing from
Eq.~\asexp\ and, therefore, from Eq.~\scheme\ for the correlation
functions. It follows immediately that {\it all the \opf\ vanish to
all orders in $g_{\rm st}$}.
In fact, all the $2k\to 1$ amplitudes vanish non-perturbatively.
To prove this, consider formula \genampl\ for a general
$n\to 1$ amplitude and perform the substitution
$x\to q-x$ in each of the integrals. Since the integrand
$r_{q-x}\,r_x$ is symmetric under this substitution, it
easily follows that
$$A_{n\to 1} = (-1)^{n+1}\,A_{n\to 1}\,\,,\eqno\eq$$
and therefore $A_{2k\to 1}=0$. This result depends
on $R_q$ being real, up to a $q$-independent overall phase.
For $\mu\neq 0$ this property is lost, so that the \opf\ no longer
vanish non-perturbatively [\rDeRo]. For $\mu=0$, on the other hand,
we expect that the \opf\ with all kinematical structures vanish
non-perturbatively.

Contrary to the \opf, the even-point functions do not vanish
and have non-trivial loop expansions. Using Eqs.~\genampl\ and
\asexp, we find for the
2--point function
$$\eqalign{&A_{1\to 1} (q,-q)=\int_0^q \,dx\,r_x r_{q-x}=
q+ {1\over 24 M}\, q^2 (7-2q^2)\cr &
+{1\over 5760 M^2} \,q^2(q-2)(501+128q-236q^2-48q^3+24 q^4 )\cr
&+{1\over 2903040 M^3}\, q^2(q-2)(q-4) (115173+67968q-49490 q^2 \cr
&\qquad\qquad\qquad\quad
-29328 q^3 +6088 q^4 +2688 q^5-464 q^6 )+\ldots
\cr }\eqn\twole$$
The one-loop result agrees with the collective field theory calculation
[\rDeRo]. The form of the higher-loop corrections is so intricate, however,
that they would be virtually impossible to obtain in the bosonized
formalism. Our results, on the other hand, give the entire non-perturbative
answer in one compact formula. We also used Eq.~\genampl\
to find the
expansion of the $3\to 1$ amplitude,
$$ \eqalign{&A_{3\to 1} (q_1,q_2,q_3;-q) =2\int_0^q \,dx\,r_x r_{q-x}
-2\sum_{i=1}^3 \int_0^{q_i} \,dx\,r_x r_{q-x}\cr
&={1\over M}\,q_1 q_2 q_3 q\, \Bigl[\,1+{1\over 24 M}\,(q-2)
\Bigl(15+4q-q^2-3(q_1^2+q_2^2+q_3^2)\Bigr)+\ldots \Bigr]\,\,.\cr }
\eqn\threeonele$$
For the $2\to 2$ kinematic structure we find, using Eq.~\twotwo,
$$\eqalign{
&A_{2\to 2} (q_1,q_2;-q_3,-q_4) = {1\over M}\,q_1 q_2 q_3 q_4\,
\Bigl[\,1+{1\over 24M}\,\Bigl(-30 + 7(q_1+q_2)\cr
\noalign{\vskip 0.2cm}
&+ 12(q_1+q_2)^2 -12(q_1 q_2+q_3 q_4) -2(q_1+q_2)^3 -
2q_1^3+ 6q_1 q_3 q_4\,\Bigr) +\ldots\Bigr]\,\,.}
\eqn\twotwole$$
Here the tree level answers agree with the collective field theory
calculations of Ref.~[\rJeYo], but the loop corrections are new.

The correlation functions
we derived constitute the Euclidean continuation of the
$S$-matrix elements of the collective field theory. In order to continue
back to the Minkowski signature, we have to take
$|q_i|\to -i\omega_i$ [\rI, \rMPR]. For instance, the $3\to 1$ amplitude
becomes
$${\cal A}_{3\to 1}(\omega_1, \omega_2, \omega_3; \omega)=
{1\over M}\,\omega_1 \omega_2 \omega_3 \omega\,
\Bigl[\,1-{1\over 24 M}\,(2+i\omega)
\Bigl(15-4i\omega+\omega^2+3(\omega_1^2+
\omega_2^2+\omega_3^2)\Bigr)+\ldots \Bigr]\,\,,
$$
where all energies are assumed positive.
This continuation takes $R_q$ into a pure phase and, according to
the arguments of Ref.~[\rMPR], the non-perturbative $S$-matrix is unitary.
This is, of course, related to the total reflection from the potential
which approaches $\infty$ as $x\to 0$.

For the Euclidean signature, the correlation functions of
the ``tachyon operators'' of string theory are simply related to
$A(q_i)$ [\rGrKl,\rDiFrKu],
$$ \VEV{T_{q_1} T_{q_2} \ldots T_{q_n}}=
A(q_1, q_2, \ldots, q_n) \,\prod_{i=1}^n L(q_i)\,\,.\eqno\eq$$
The external leg factor $L(q)$ can be calculated
from a matrix model representation of the tachyon operator [\rMPR],
$$ T_q \sim f(|q|) \int dt \,\,e^{iqt}\, \Tr\, e^{-l \Phi^2 (t)}\,\,,
\eqno\eq$$
where $f(|q|)$ is a smooth function which determines the normalization.
One finds that
$$ L(q) \sim \int^\infty d\tau e^{-l \sqrt M \cosh 2\tau} e^{-|q|\tau}
\sim (l\sqrt M)^{|q|/2} \Gamma (-q|/2)\,\,.\eqno\eq$$
We may chose the operator normalization $f(|q|)$ so that
$$L(q) =M^{|q|/4}\,\,{\Gamma (-|q|/2)\over \Gamma (|q|/2)}\,\,.\eqno\eq$$
Now $ L(i\omega)$ is a pure phase, as needed for the
unitarity of the Minkowski signature $S$-matrix.
Note that $L(q)$ has poles for $|q|= 2n$, $n>0$, while for
the conventional $c=1$ model the poles occur for $|q|=n>0$.
This agrees with an argument for the position of the poles
based on energy sum rules in the black hole conformal field
theory [\rJeYo,\rEg]. Reproducing our exact correlation functions in the
context of conformal field theory poses an interesting challenge.

In this Letter we calculated the exact $S$-matrix of the deformed
$c=1$ \mm\ for $M>0$, which has been conjectured to describe
the stationary black hole background of \td\ string theory.
Even if the black hole analogy fails, this model is interesting in its
own right because it leads to a new non-perturbatively calculable
unitary $S$-matrix.
There are many interesting extensions of this work.
For example, one may consider the case of $M<0$, which
has been conjectured to describe a
``naked singularity'' [\rJeYo].
We hope to return to these problems in a future paper.

\ack
We wish to thank Antal Jevicki for interesting discussions.
I. R. K. and K. D. are supported in part by
the NSF Presidential Young Investigator Award PHY-9157482 and
James S. McDonnell Foundation grant No. 91-48. I. R. K. is also
supported in part by DOE grant DE-AC02-76WRO3072 and
an A. P. Sloan Foundation Research Fellowship.

\singlespace
\refout
\bye